\documentclass[a4paper,11pt]{article}
\pdfoutput=1 

\usepackage{jheppub} 

\usepackage{graphicx,epsfig,color}
\usepackage[english]{babel}
\usepackage{amssymb,amsfonts,amsmath}
\usepackage{verbatim}
\usepackage{bbm,bbold}
\usepackage{bigints}
  \newcommand{\cB}{{\cal B}}

  \newcommand{\cL}{{\cal L}}
\newcommand{\cM}{{\cal M}}

\newcommand{\cS}{{\cal S}}

\newcommand{\be}{\begin{equation}} \newcommand{\ee}{\end{equation}}
\newcommand{\bea}{\begin{eqnarray}} \newcommand{\eea}{\end{eqnarray}}
\newcommand{\beann}{\begin{eqnarray*}}  \newcommand{\eeann}{\end{eqnarray*}}
\newcommand{\bfig}{\begin{figure}} \newcommand{\efig}{\end{figure}}
\newcommand{\ba}{\begin{array}} \newcommand{\ea}{\end{array}}
\newcommand{\bcen}{\begin{center}} \newcommand{\ecen}{\end{center}}
\newcommand{\btab}{\begin{tabular}} \newcommand{\etab}{\end{tabular}}

\newcommand{\e}{{\rm e}}

\newtheorem{Proposition}{Proposition}[section]

\newtheorem{Theorem}{Theorem}[section]
\newtheorem{Lemma}{Lemma}[section]
\newtheorem{Corrolary}{Corrolary}[section]

\newcommand{\bp}{\begin{Proposition}}   \newcommand{\ep}{\end{Proposition}}
\newcommand{\bt}{\begin{Theorem}}   \newcommand{\et}{\end{Theorem}}
\newcommand{\bl}{\begin{Lemma}}     \newcommand{\el}{\end{Lemma}}
\newcommand{\bc}{\begin{Corrolary}} \newcommand{\ec}{\end{Corrolary}}
\title{\boldmath Hall Viscosity in a Strongly Coupled Magnetized Plasma}

\author[a,b]{Carlos Hoyos,}
\author[c]{Francisco Pe\~na-Benitez,}
\author[c]{Piotr Witkowski}

\affiliation[a]{Dept. of Physics, Universidad de Oviedo, Federico Garc\'ia Lorca 18, 33007, Oviedo, Spain}
\affiliation[b]{ICTEA, Calle de la Independencia, 13, 33004 Oviedo, Spain}
\affiliation[c]{Max-Planck-Institut  f\"ur Physik komplexer Systeme, N\"othnitzer Str. 38, 01187 Dresden, Germany}

\emailAdd{hoyoscarlos@unovi.es}
\emailAdd{pena@pks.mpg.de}
\emailAdd{piotr@pks.mpg.de}

\abstract{We show how a Hall viscosity induced by a magnetic field can be generated in strongly coupled theories with a holographic dual. This is achieved by considering parity-breaking higher derivative terms in the gravity dual. These terms couple the Riemann curvature tensor to the field strength of a gauge field dual to the charge current, and have an analog in the field theory side as a coupling between the ``Euler current'' and the electromagnetic field. As a concrete example, we study the effect of the new terms in the thermodynamic and transport properties of a strongly coupled magnetized plasma dual to a dyonic black hole in $AdS_4$. As a new property of the holographic model, we find that for a state that is initially neutral at zero magnetic field, a charge density and non-dissipative Hall transport are present when the magnetic field is turned on. Remarkably, we also observe that the results from the holographic model are consistent with hydrodynamics even at magnetic fields much larger than temperature.}

\begin{document}

\maketitle

\section{Introduction and summary}

Some of the most fascinating systems in condensed matter are the Quantum Hall (QH) states. They are typically characterized by the value of the Hall conductivity, which is a topologically protected quantity proportional to an integer (IQH) or a fraction (FQH) times a quantum unit of  conductance.

Although the IQH states can be understood in terms of free electrons in completely filled Landau levels, the FQH states are a consequence of strong correlations between electrons in partially filled Landau levels. There are quite successful phenomenological descriptions of FQH states in terms of Laughlin wavefunctions 
\cite{PhysRevLett.50.1395} or composite fermions, effective bound states of fermions with magnetic fluxes \cite{PhysRevLett.63.199}.\footnote{Recently, particle-hole symmetry of the underlying microscopic theory has been used to argue that composite fermions should be relativistic \cite{Son:2015xqa}, although this is still under discussion \cite{Wang:2017cmz,Nguyen:2017mcw}.}  However, barring numerical simulations, we still lack a first principles derivation of the properties of FQH states. In some aspects, the situation mirrors quantum chromodynamics (QCD) in high energy physics. 

In FQH states, the Hall conductivity is not the only topologically protected quantity. If the fermions are put on a sphere, the ratio between the number of particles $N_e$ and flux quanta $N_\Phi$ is not anymore determined by the filling fraction $\nu$, but there is a shift $\cS=\nu^{-1}N_e- N_\Phi$, which is a fractional number and can be used to further characterize the state \cite{PhysRevLett.69.953}. If the fermions are in flat space and rotational invariance is a good symmetry on the average, the shift determines the value of the Hall viscosity \cite{PhysRevLett.75.697,1997physics..12050A,PhysRevB.79.045308} -- a transport coefficient analogous to the Hall conductivity for external strain rate instead of electric field. Similarly to the Hall conductivity, the Hall viscosity can only be nonzero if time reversal invariance and parity are broken, as it is the case when a background magnetic field is present. This will also be the case for compressible states like two-dimensional metals, where Hall viscosity can produce measurable effects in Hall transport in a hydrodynamic regime \cite{Scaffidi2017,narozhny2019magnetohydrodynamics,berdyugin2019measuring,Matthaiakakis:2019swi}.

A deeper understanding of FQH states and other strongly correlated electron states  in condensed matter may be gained by studying other strongly coupled systems exhibiting similar properties. Typically they would be as hard to study as the original problem but, thanks to the AdS/CFT correspondence (or holographic duality) \cite{Maldacena:1997re,Gubser:1998bc,Witten:1998qj}, a whole class of strongly coupled theories can be studied  using much simpler classical theories of gravity. Although theories with holographic duals can be microscopically very different to a typical condensed system, their long wavelength behavior can be similar and be used as a playing ground to understand better the properties of FQH states. Indeed, there have been many proposals at describing QH states using holography  \cite{KeskiVakkuri:2008eb,Davis:2008nv,Fujita:2009kw,Hikida:2009tp,Kawamoto:2009sn,Belhaj:2010nn,Bergman:2010gm,Goldstein:2010aw,Bayntun:2010nx,Fujita:2010pj,Jokela:2011eb,Fujita:2012fp,Blake:2012tp,Melnikov:2012tb,Kristjansen:2012ny,Jokela:2013hta,Lippert:2014jma,Bea:2014yda,Lindgren:2015lia,Mezzalira:2015vzn}. Although some of these models describe fractional filling fractions and features like transitions between Hall plateaus, the situation is not entirely satisfactory when the holographic models are examined in detail. In most cases the Hall conductivity is non-vanishing even in the absence of a magnetic field, thus these models have an ``anomalous'' Hall effect, rather than the usual Hall effect induced by a magnetic field. From the point of view of the holographic dual, the Hall conductivity is generated by a topological term of the form
\be\label{eq:FwedgeF}
S_{\rm top}=\sigma_H\int F\wedge F.
\ee
In general, a Hall conductivity generated by this kind of term does not need to be quantized, although in particular models it can be. A second issue, that will be the focus of this paper, is that the value of the Hall viscosity in existing holographic models of QH states is likely to vanish, although this question has not been studied systematically.

Why would the Hall viscosity vanish in holographic models of QH states? The simplest examples of holographic models with non-zero Hall conductivity and zero Hall viscosity are four-dimensional dyonic black hole solutions of Einstein-Maxwell theory \cite{Hartnoll:2007ai}. Refinements with axion and dilaton fields \cite{Goldstein:2010aw,Bayntun:2010nx,Lindgren:2015lia} modify the scalar sector, so in principle they would not affect the viscosity, which is part of the tensor components of stress. A non-zero Hall viscosity could appear if an axion is coupled to a $R\wedge R$ term \cite{Saremi:2011ab},\footnote{There are other ways to generate a Hall viscosity in holographic models of chiral superfluids, see e.g. \cite{Son:2013xra,Hoyos:2014nua}.} but in this case the Hall viscosity is not produced by the magnetic field, but by a pseudoscalar mass or condensate, in analogy to the case of the Haldane model \cite{Haldanemodel}, where a time reversal breaking mass allows for Hall transport in absence of magnetic fields. 

A pertinent question in this program is then what are the necessary modifications of holographic models such that the magnetic field will produce a non-zero Hall viscosity. In order to find an answer, one can look for inspiration in the effective long wave description of QH states. These are gapped systems, so that, at low enough frequencies and wave number, the response to external fields is described by a local functional of the external sources, in this case an electromagnetic field $A_\mu$ and the metric $g_{\mu\nu}$ (or in non-relativistic system, the spatial metric $g_{ij}$). The term responsible for the Hall conductivity is a Chern-Simons action for the external electromagnetic field
\be
S_{CS}=\frac{\nu}{2\pi}\int A\wedge dA.
\ee 
For the Hall viscosity in a non-relativistic system, there is a mixed Chern-Simons known as the Wen-Zee term \cite{PhysRevLett.69.953}
\be
S_{WZ}=\frac{\nu\cS}{16\pi}\int A\wedge R,
\ee
where $R$ is the curvature two-form associated to the spatial metric.  In a relativistic system, it is not possible to construct a covariant Chern-Simons term mixing the electromagnetic field with the metric. There is nevertheless a covariant term that reduces to the Wen-Zee term in the non-relativistic limit obtained by taking the speed of light to infinity $c\to \infty$ and that is constructed with an algebraically conserved Euler current \cite{Golkar:2014wwa,Golkar:2014paa}
\be
S_{WZ\,{\rm rel}}= \kappa \int \sqrt{-g} \, A_\mu J^\mu_E,
\ee
where
\be
J_E^\mu=\frac{1}{8\pi}\varepsilon^{\mu\nu\rho}\varepsilon^{\alpha\beta\gamma} u_\alpha\left( \nabla_\nu u_\beta\nabla_\rho u_\gamma-\frac{1}{2}R_{\nu\rho\beta\gamma}\right).
\ee
The four velocity $u^\mu u_\mu=-1$ is defined using the external electromagnetic field
\be
u^\mu=\frac{1}{2b}\varepsilon^{\mu\nu\lambda} F_{\nu\lambda},\ \ b^2=\frac{1}{2} F_{\mu\nu}F^{\mu\nu}.
\ee
Clearly, an effective action with this term can only be well defined if the background magnetic field is nonzero. Expanding around a constant magnetic field $F_{0i}=\delta F_{0i}$, $F_{ij}=\varepsilon_{ij}B+\delta F_{ij}$, 
\be
S_{WZ\,{\rm rel}}\sim  \kappa \int \sqrt{-g} \, \varepsilon^{\mu\nu\lambda}A_\mu R_{\nu\lambda i j}\varepsilon^{ij}.
\ee
For $\kappa\sim B$, the term above resembles a boundary term 
\be \label{eq:SbdrStruct}
\begin{split}
&S_{WZ\,{\rm rel}}\sim   \int_{\partial \cM} d^3 x \, \epsilon^{\mu\nu\lambda} \delta A_\mu R_{\nu\lambda i j}F^{ij}\\
& \sim \int_{\cM} d^4 x \, \partial_4\left(  \epsilon^{4\mu\nu\lambda} \delta A_\mu R_{\nu\lambda i j}F^{ij} \right) \sim  \int_{\cM} d^4 x\, \epsilon^{4\mu\nu\lambda} \delta  F_{4\mu} R_{\nu\lambda i j}F^{ij}.
\end{split}
\ee
Where $\cM$ would be some four-dimensional manifold and $\partial\cM$ its boundary. Although the formulas above are not exact, they suggest that the Hall viscosity in a holographic model will be produced by higher derivative terms in the four-dimensional action with a similar structure. We will show that this is indeed the case and find the transport coefficients as a function of the magnetic field and temperature using a perturbative expansion in the coefficients of the higher derivative terms. 

For simplicity, our starting point will be the dyonic black hole solutions of Einstein-Maxwell theory mentioned previously. They are dual to a family of states characterized by their temperature $T$, chemical potential $\mu$ and magnetic field $B$, and belonging to a strongly coupled theory (typically a gauge theory in a large-$N$ limit). The conductivity matrix for charge $\vec{J}$ and heat $\vec{Q}$ currents is  usually defined as
\be
\left( \begin{array}{c} \vec{J} \\  \vec{Q} \end{array}\right)=\left( 
\begin{array}{cc}
\sigma & \alpha \\
T \alpha & \bar{\kappa}
\end{array}
\right)\left( \begin{array}{c} \vec{E} \\  -\vec{\nabla}T \end{array}\right),
\ee
where $\vec{E}$ is the electric field and $\vec{\nabla} T$ the temperature gradient. The Hall component of the conductivities is the antisymmetric part of the matrices $\sigma$, $\alpha$ and $\bar{\kappa}$, that are identified with charge, thermoelectric and heat conductivities respectively. At zero chemical potential the charge density $\rho$ vanishes and as a consequence the Hall conductivity  is zero $\sigma_{_H}=0$. The Hall heat conductivity also vanishes at zero chemical potential $\bar{\kappa}_{_H}=0$, although the thermoelectric Hall conductivity does not $\alpha_{_H}\neq 0$.

After introducing higher derivative terms inspired by \eqref{eq:SbdrStruct}, it turns out that the charge density no longer vanishes, but it is proportional to the magnetic field, with a coefficient that depends on the dimensionless ratio $B/T^2$. We confirm that indeed the Hall viscosity is non-zero and proportional to the magnetic field $\eta_{_H}\propto B$, our main result. We also have checked that our results for the transport coefficients are consistent with hydrodynamics in the presence of an external magnetic field. The Hall conductivity is equal to $\rho/B$ as expected, and the heat thermal and thermoelectric  Hall conductivities are non-zero. All the transport coefficients have factors that depend on $B/T^2$ and the coefficients of the higher derivative terms, in particular, the Hall viscosity shows a different power dependence at low and high values of the temperature
\be
\eta_{_H} \sim \left\{\begin{array}{lr}
	B^2/T^2\,, & B/T^2\ll 1\\
	B\,, & B/T^2 \gg 1
	\end{array}\right.\,.
\ee
Interestingly, by including a term of the form \eqref{eq:FwedgeF}, such that the anomalous Hall conductivity is tuned to vanish, all the odd transport coefficients are proportional to the same combination of coefficients of the higher derivative terms.  In this case, at large magnetic fields, the Hall viscosity becomes proportional to the charge density
\be
\eta_{_H} \simeq \frac{5}{\sqrt{3}}\rho+\mathcal O(T/\sqrt{B}).
\ee

The paper is organized as follows, in Sect. \ref{sec:hydro} we review the hydrodynamic description of a parity-breaking fluid, and obtain the proper Kubo formulas relating the transport coefficients of the system with the two point functions of the corresponding conserved currents in the presence of a homogeneous magnetic field. In Sect. \ref{sec:model} we define the holographic model, write the equations of motion and compute the holographic one-point functions of the theory. Then, we construct the deformed dyonic black hole and analyse its thermodynamic properties. To continue with the study of our holographic model, we analytically compute all the transport coefficients of the system in Sect. \ref{sec:transp} and compare with the results from hydrodynamics. Finally, in Sect. \ref{sec:discussion} we discuss our results and the low and high temperature limits and comment on possible outlooks.  We also accompany the paper with a series of appendices where we discuss the variational problem and renormalization of the theory (see appendix \ref{app:LCT}), and collect lengthy formulas such as the equations of motion (appendix \ref{app:eqs}), their perturbative black holes solution (\ref{app:background}), or the sources (appendix \ref{app:sources}) appearing in the equations of motion for the vector sector.

\section{Hydrodynamics with a magnetic field}\label{sec:hydro}

The low energy effective description of the holographic dual to a black hole is determined by relativistic fluid equations \cite{Bhattacharyya:2008jc}, for a charged fluid
\be\label{eq:eqshydro}
\nabla_\alpha T^{\alpha\mu}=F^{\mu\beta}J_\beta,\ \ \nabla_\mu J^\mu=0,
\ee
where $T^{\mu\nu}$ is the energy-momentum tensor and $J^\mu$ the charge current. In the case we want to study a background magnetic field has been turned on, so that parity is broken. The general form of the constitutive relations for this type of fluid in $2+1$ dimensions has been worked out in \cite{Jensen:2011xb}. Although in their case the magnetic field was constrained to be much smaller than the temperature, we will assume that the constitutive relations can be generalized to larger values of the magnetic field by allowing thermodynamic potentials to depend on it in the frame referred to as ``magnetovortical''. In principle there could be terms that depend on gradients of the magnetic field, but as we will restrict to a homogeneous and constant background, those can be safely ignored. We will partially confirm the validity of the assumption above by deriving linear response results using hydrodynamic equations and comparing with the holographic model. Some of the linear response coefficients are determined by thermodynamic quantities, and those should match, while the others will be used to fix the value of transport coefficients in the hydrodynamic constitutive relations. 

We proceed to present the constitutive relations. The hydrodynamic variables are the fluid velocity $u^\mu$, $u^\mu u_\mu=-1$, temperature $T$ and chemical potential $\mu$. The energy-momentum and current are expanded in derivatives of those and the background sources. At zeroth order one finds the pressure $P$, energy density $\varepsilon$ and magnetization $M$, all of which are functions of $\mu$ $T$ and the background magnetic field $B$. In order to have covariant expressions, the magnetic field dependence will enter through the pseudoscalar
\be
\cB=-\frac{1}{2}\varepsilon^{\mu\nu\lambda} u_\mu F_{\nu\lambda},
\ee
where $F_{\mu\nu}$ is the field strength of the background gauge field $A_\mu$. The  curly epsilon symbol is defined as a tensor, i.e with a factor of the metric determinant $\varepsilon^{012}=+\frac{1}{\sqrt{-g}}$. With these ingredients, the constitutive relations are
\be\label{eq:constrel}
T^{\mu\nu}=\varepsilon u^\mu u^\nu+(P-\cB M) P^{\mu\nu}+\tau^{\mu\nu},\ \ J^\mu=\rho u^\mu+\nu^\mu.
\ee
Here $P ^{\mu\nu}=g^{\mu\nu}+u^\mu u^\nu$ is the projector transverse to the velocity and $\tau^{\mu\nu}$, $\nu^\mu$ contain higher derivative terms and satisfy the condition
\be
\nu^\mu u_\mu=0,\ \tau^{\mu\nu}u_\nu=0.
\ee
The magnetization should satisfy
\be
M=\left(\frac{\partial P}{\partial \cB}\right)_{T,\mu}.
\ee
The energy density, pressure, entropy density and charge satisfy the thermodynamic relations
\be
\varepsilon+P=Ts+\mu \rho,\ \ s=\left(\frac{\partial P}{\partial T}\right)_{\mu,\cB},\ \rho=\left(\frac{\partial P}{\partial \mu}\right)_{T,\cB}.
\ee
At first order in the derivative expansion, the independent contributions allowed by the second law condition are \cite{Jensen:2011xb}\footnote{We have dropped terms proportional to the vorticity, as those will not play a role in our analysis, in addition we have introduced $\bar\sigma_{_{H}}$, which is related with $\tilde\chi_{_E}$ of \cite{Jensen:2011xb} as $\bar\sigma_{_{H}}=\tilde\sigma_{_{V}}+\tilde\chi_{_E}$.}
\be
\begin{split}
&\tau^{\mu\nu}=-\eta \Sigma^{\mu\nu}-\eta_{_H}\tilde \Sigma^{\mu\nu}-\zeta P ^{\mu\nu}\nabla_\alpha u^\alpha,\\
&\nu^\mu=\sigma_{_V} V^\mu+\bar\sigma_{_H} \tilde{E}^\mu-T\tilde\sigma_{_V} \varepsilon^{\mu\nu\lambda}u_\nu \nabla_\lambda \frac{\mu}{T}+\tilde{\chi}_T \varepsilon^{\mu\nu\lambda}u_\nu \nabla_\lambda T.
\end{split}
\ee 
The expressions for each of the objects appearing in the formulas above are
\be
\begin{split}
&\Sigma^{\mu\nu}=P ^{\mu\alpha}P^{\nu\beta}\left(\nabla_\alpha u_\beta+\nabla_\beta u_\alpha-g_{\alpha\beta}\nabla_\sigma u ^\sigma\right),\\
&V^\mu=E^\mu-TP ^{\mu\alpha}\nabla_\alpha\frac{\mu}{T},\ \ E^\mu= F^{\mu\alpha} u_\alpha,\\
&   \tilde{E}^\mu=\varepsilon^{\mu\nu\lambda}u_\nu E_\lambda,\ \
\tilde{\Sigma}^{\mu\nu}=\varepsilon^{(\mu|\alpha\beta}u_\alpha \Sigma_\beta^{\ \nu)}.
\end{split}
\ee

This completes our setup. For a flat background and constant magnetic field $\cB=B$ the energy-momentum tensor and current are
\be
T^{00}=\varepsilon, \ \ T^{ij}=(P-BM)\delta^{ij},\ \ J^0=\rho,\ \ J^i=0.
\ee
In order to extract the transport properties, we will perturb the metric and the gauge field and solve the equations \eqref{eq:eqshydro} for $u^\mu$, $T$ and $\mu$ to linear order in the perturbations. We then will evaluate the energy-momentum tensor and the current on the solutions. The two-point functions are obtained by taking variations with respect to the sources of the energy-momentum tensor and the current
\be
\begin{split}
&G_{TT}^{\mu\nu,\alpha\beta}=\frac{\delta T^{\mu\nu}}{\delta g_{\alpha\beta}},\ \ G_{JJ}^{\mu,\nu}=\frac{\delta J^{\mu}}{\delta A_\nu},\\
& G_{TJ}^{\mu\nu,\lambda}=\frac{\delta T^{\mu\nu}}{\delta A_\lambda}, \ \ G_{JT}^{\lambda,\mu\nu}=\frac{\delta J^{\lambda}}{\delta g_{\mu\nu}}.
\end{split}
\ee
For simplicity we will consider only homogeneous but time-dependent sources, the metric and gauge field will be 
\be
g_{\mu\nu}=\eta_{\mu\nu}+\epsilon\, h_{\mu\nu}(t),\ \ A=\epsilon\, a(t)dt+\left(-\frac{B}{2} \epsilon_{ij}x^j+\epsilon\, a_i(t)\right)dx^i.
\ee
The parameter $\epsilon \ll 1$ is small, so that we are in the regime of linear response for the perturbations $h_{\mu\nu}$ and $a_\mu$. The perturbations will be expanded in plane waves
\be
h_{\mu\nu}(t)=\int \frac{d\omega}{2\pi} \hat{h}_{\mu\nu}(\omega) e^{-i\omega t}, \ \ a_\mu(t)=\int \frac{d\omega}{2\pi} \hat{a}_\mu(\omega) e^{-i\omega t},
\ee
and only low frequency components will be nonzero, in such a way that the equations for each Fourier component are expanded up to linear order in $\omega$ and higher orders are neglected. When solving the equations different sectors decouple according to their representation under spatial rotations. We can distinguish a tensor, vector and scalar sector related respectively to shear and Hall viscosities, conductivities and bulk viscosity.

 After substituting Eqs. \eqref{eq:constrel} into the conservation equations \eqref{eq:eqshydro} and solving for the velocities, temperature and chemical potential the results are, in the tensor sector,
\be
\label{eq:TensorKubo} G_{TT}^{xy,xy}=\frac{1}{4}G_{TT}^{xx-yy,xx-yy}=-(P-B M)+i\omega \eta,\ \ G_{TT}^{xy,xx-yy}=-G_{TT}^{xx-yy,xy}=-i\eta_{_H}\omega
\ee
The scalar sector contains a term linear in the frequency proportional to the bulk viscosity $\zeta$
\be
\frac{1}{4}G_{TT}^{xx+yy,xx+yy}\supset i\zeta\omega.
\ee

There is also a zero frequency contribution that is a fairly complicated combination of thermodynamic derivatives. This contribution is analogous to the inverse compressibility term appearing in non-relativistic systems \cite{Bradlyn:2012ea}. At zero magnetic field the structure of the correlators does not change, one can use the expression above setting $B=0$.

In the vector sector the current-current and current-momentum correlators are
\be\label{eq:JJJT}
G_{JJ}^{ij}=i\frac{\rho}{B}\omega \epsilon^{ij},\ \ G_{TJ}^{0i,j}=G_{JT}^{i,0j}=i\frac{\varepsilon+P-B M}{B}\omega \epsilon^{ij}.
\ee
The momentum-momentum correlator is
\be\label{eq:GTT}
G_{TT}^{0i,0j}=\left(P-B M+i K \sigma_V \omega \right)\delta^{ij}+iK\left(\frac{\rho}{B}-\bar\sigma_{_{H}}\right) \omega \epsilon^{ij},
\ee
where
\be\label{eq:K}
K=\frac{(\varepsilon +P- B M)^2}{B^2\sigma_{_V}^2+\left( \rho-B\bar\sigma_{_{H}}\right)^2}\,,
\ee
which can be used to compute the transport coefficients $\sigma_{_{V}}$ and $\bar\sigma_{_{H}}$. Notice that the susceptibilities $\tilde\chi_{_E}=\bar\sigma_{_H}-\tilde\sigma$ and $\tilde\chi_{_T}$ can be computed combining the zero frequency correlators $G_{JJ}^{x,t}(0,q_y)$ and  $G_{JT}^{x,tt}(0,q_y)$, which in the hydrodynamic regime read
 \begin{align}
 \lim_{q_y \to 0}\frac{1}{q_y}{\rm Im}G_{JJ}^{1,0}(0,q_y) &=   \left(\frac{ \tilde\chi_{_T} (\partial M/\partial\mu-\tilde\chi_{_E})T\,B}{ \mu  \left(\partial M/\partial\mu\right) B+B T (\partial M/\partial T+ \tilde\chi_{_T})-\varepsilon -P}+\tilde\chi_{_E}\right)\,,\\
\lim_{q_y \to 0} \frac{1}{q_y}{\rm Im}G_{JT}^{1,00}(0,q_y) &=  \frac{1}{2} \frac{ T \tilde\chi_{_T} (\varepsilon+P-M B)}{ \mu \left(\partial M/\partial\mu\right) B+B T (\partial M/\partial T+ \tilde\chi_{_T})-\varepsilon -P}\,,
 \end{align}
however, we have not explicitly computed these quantities for the holographic system because they do not contribute to the actual transport, as we show below. In fact, 
following \cite{Hartnoll:2007ih}, the heat current is defined as a combination of energy and charge currents $Q^i= T^{0i}-\mu J^i$,\footnote{Recall that in a relativistic system energy current is the same as momentum density.} and the conductivities correspond to the linear response of the charge and heat current to electric fields and temperature gradients
\be
\left( \begin{array}{c} \vec{J} \\  \vec{Q} \end{array}\right)=\left( 
\begin{array}{cc}
\sigma & \alpha \\
T \alpha & \bar{\kappa}
\end{array}
\right)\left( \begin{array}{c} \vec{E} \\  -\vec{\nabla}T \end{array}\right)\,.
\ee
The conductivities are then defined through the Kubo formulas.\footnote{We use a different sign in the Kubo formulas because of our conventions. The electric field has been defined as $E_i=F_{i0} u^0=\partial_i A-\partial_t A_i$. Then, Ohm's law becames $J^i=\sigma^{ij} E_j\simeq -\sigma^{ij}\partial_t A_j \sim +i \omega \sigma^{ij}A_j$. Hence the conductivity is extracted as proportional to the positive imaginary part of the correlator. More generally, with our conventions the spectral function is proportional to the imaginary part of the retarded correlator obtained through the linear response formulas.}
\be
\begin{split}
&\sigma^{ij}=\lim_{\omega \to 0} \frac{1}{\omega}{\rm Im}\,G_{JJ}^{ij},\\
&T\alpha^{ij}=\lim_{\omega \to 0} \frac{1}{\omega}{\rm Im}\,\left( G_{JT}^{i,0j}-\mu G_{JJ}^{ij}\right)+M \epsilon^{ij},\\
&T\bar{\kappa}^{ij}=\lim_{\omega\to 0}\frac{1}{\omega}{\rm Im}\, \left(G_{TT}^{0i,0j}-\mu G_{TJ}^{0i,j}-\mu G_{JT}^{i,0j}+\mu^2 G_{JJ}^{ij} \right) -2\mu M\epsilon^{ij}.\label{eq:Kubo}
\end{split}
\ee
The last terms in the Kubo formulas for thermoelectric ($\alpha$) and heat ($\bar{\kappa}$) conductivities subtract the contributions from the magnetization current, which are not part of the transport by motion of charge carriers. We have checked that our results agree with \cite{Hartnoll:2007ai,Hartnoll:2007ih}. These coefficients define the response in the absence of either electric field or temperature gradient. It is customary to study heat transport in the absence of electric current, for which one can define the thermal conductivity $\kappa$ as the heat current produced by a temperature gradient, and thermoelectric response as the electric field in this situation $\vec{E}=-\vartheta \vec{\nabla}T$.  The response coefficients are determined by
\be
\kappa=\bar{\kappa}-T \alpha \sigma^{-1}\alpha,\ \ \vartheta=-\sigma^{-1}\alpha.
\ee
Using the results obtained from the hydrodynamic equations, the values of the electric and thermoelectric conductivities are determined by the charge and entropy densities
\be
\sigma^{ij}=\frac{\rho}{B}\epsilon^{ij},\ \ \alpha^{ij}=\frac{s}{B}\epsilon^{ij}.\label{eq:ConductivitiesHydro}
\ee
The heat conductivity (in the absence of electric fields) is, for $B/T^2\ll 1$,
\be
\bar{\kappa}^{ij}=\frac{(\varepsilon+P)^2}{T\rho^2}\sigma_{_V} \delta^{ij}+\left(\frac{T s^2}{\rho B}+\frac{(\varepsilon+P)^2}{T\rho^2}\bar\sigma_{_{H}}\right)\epsilon^{ij}.
\ee
While the thermal conductivity (for $B/T^2\ll 1$) and Seebeck coefficients are
\be
\kappa^{ij}=\frac{(\varepsilon+P)^2}{T\rho^2}\left(\sigma_{_V} \delta^{ij}+\bar\sigma_{_{H}}\epsilon^{ij}\right), \ \ \vartheta^{ij}=-\frac{s}{\rho}\delta^{ij}.
\ee
Therefore, the terms appearing in the constitutive relations of the charge current manifest themselves as contributions to the thermal conductivity, in particular the parity odd coefficient $\bar\sigma_{_{H}}$,  induces a thermal Hall conductivity that would otherwise be vanishing. 
 
At zero magnetic field the structure of the vector correlators and conductivities changes significantly. The current-current  correlator becomes 
\be
G_{JJ}^{ij}=\left(-\frac{\rho^2}{\varepsilon+P}+i\sigma_{_V}\omega\right) \delta^{ij}+i \bar\sigma_{_{H}}\omega \epsilon^{ij}.
\ee
While the current-momentum and momentum-momentum correlator take the simple form
\be
G_{TJ}^{0i,j}=G_{JT}^{i,0j}=-\rho \delta^{ij},\ \ G_{TT}^{0i,0j}=-\varepsilon \delta^{ij}.
\ee
Then, the conductivities at zero chemical potential are (assuming the magnetization vanishes as well)
\be\label{eq:condBzero}
\sigma^{ij}=\sigma_{_V}\delta^{ij} +\bar\sigma_{_H}\epsilon^{ij},\ \ \alpha^{ij}=\bar{\kappa}^{ij}=0.
\ee
In this case $\sigma_{_V}$ is the longitudinal conductivity, while the combination $\bar\sigma_{_H}$ enters as an (anomalous) Hall conductivity.   At zero chemical potential the charged current in the system is produced by particle-antiparticles pair. The flow of those particles transport a net charge, but the total momentum flow is zero, because particles and antiparticles move in opposite directions. This fact explains the vanishing value for both the thermoelectric and thermal conductivities.

\section{Holographic model}\label{sec:model}

The simplest holographic model that incorporating magnetic fields is the Einstein-Maxwell theory, which admits dyonic black hole solutions dual to states of a strongly coupled theory at nonzero temperature, charge density and magnetic field  \cite{Hartnoll:2007ai}. When both charge and magnetic field are present, there is a nonzero Hall conductivity, but the Hall viscosity vanishes, even though it is not forbidden by symmetries and it is generated by a magnetic field in other cases, such as Quantum Hall systems. Our goal is to identify the ingredients necessary in a holographic model such that a Hall viscosity will be induced when we apply a magnetic field. Motivated by the arguments explained in the introduction, we consider a higher derivative gravity model with extra terms breaking the parity invariance of the system, and which should have at least four derivatives. For simplicity we will ignore all the four derivative terms that are parity even and we will not consider terms with three derivatives (such as $\sim F^3$) or more than four derivatives. With these restrictions, the more general action we can write reads
\begin{equation}
S = \frac{1}{2\kappa^2}\int d^4x\sqrt{-g}\left[ R+\frac{6}{L^2} -\frac{L^2}{4}F^{MN}F_{MN} + \lambda_0 L^2\mathcal{L}_0  + \sum_{i=1}^5\lambda_i L^4\mathcal{L}_i \right]+S_{GH}+S_{CT}\,,\label{eq:Action}
\end{equation}
where
\begin{align}
&\mathcal L_0=\epsilon^{MNPQ} F_{MN} F_{PQ}\,,\qquad\qquad\mathcal{L}_1= \epsilon^{MNPQ}F_{MN}F^{A}_{~B}R^B_{~APQ},\\
&\mathcal{L}_2=\epsilon^{MNPQ}R_{SRPQ}F_M^{~S} F_N^{~R}\,,\quad\quad \mathcal{L}_3=\epsilon^{MNRP}R_{ARBP}F_M^{~A} F_N^{~B},\\
&\mathcal L_4=\epsilon^{MNPQ}\nabla_A F_{MN}\nabla^A F_{PQ}\,,\quad\quad \mathcal L_5=\frac{1}{2}\epsilon^{MNPQ} R^A\,_{BMN} R^B\,_{APQ}\,,
\end{align}
where $\epsilon^{trxy}=-\frac{1}{\sqrt{-g}}$.

The term $\mathcal{L}_0$ introduces an anomalous Hall conductivity, while the term $\mathcal{L}_5$ in principle does not affect to the first order transport coefficients. In the literature $\mathcal{L}_0$ \cite{Jensen:2011xb}, and $\mathcal{L}_5$ \cite{Saremi:2011ab} have also been considered including a  coupling  to an axion field. In this case, the last can produce a non-zero Hall viscosity. However, as the axion would count as a different source of parity breaking not related to the magnetic field, we keep the coefficients of these terms constant and drop $\cL_5$ from the subsequent analysis. On top of this, it can be shown that among the $\mathcal L_1,\ldots,\mathcal L_5$ only two of them are independent.  Therefore, we will keep only $\lambda_0,\lambda_1$, and $\lambda_3$ non-vanishing.  It is straightforward to derive the equations of motion for this action, but as they are not very illuminating, we have collected them in Appendix~\ref{app:eqs}. The most apparent change that the new terms introduce is that the electric flux is not necessarily the same at the black hole horizon and the boundary, we will comment more on this in the discussion. 

\subsection{Background solutions and thermodynamics}

In the absence of higher derivative terms $\lambda_1=\lambda_3=0$, the action \eqref{eq:Action} admits dyonic black hole solutions with non-zero charge and magnetic field. Assuming that in a consistent truncation of supergravity the couplings $\lambda_1$ and $\lambda_3$ should come as subleading corrections in the dual large-$N$, strong coupling expansions, and to avoid all the subtleties associated to having a higher derivative theory of gravity,  we will treat those parameters perturbatively, and therefore the black hole solution shall be similar to the dyonic black hole.  This allows us to start with the following ansatz that generalizes the dyonic black hole solution
\begin{equation}
\frac{ds^2}{L^2}=\frac{1}{r^2f(r)}dr^2+r_0^2 r^2\left(-f(r)dt^2+C(r)\left(dx^2+dy^2\right)\right)\,,\quad A=a(r)dt+B x dy\,,\label{eq:Ansatz}
\end{equation} 
where the factor $r_0$ comes from a convenient re-scaling of the equations such that the black hole horizon is located at $r=1$. Note that the coordinate $r$ is dimensionless, but $r_0$ has dimensions of energy.  Regularity of the Euclidean solution demands that we impose the boundary conditions
\be
f(1)=a(1)=0.
\ee
The chemical potential is then determined by the value of the gauge potential at the boundary
\be
\mu=\lim_{r\to \infty} a(r).
\ee
Considering that the solutions are lengthy and not particularly enlightening, for the general case of $\mu\neq 0$ we relegate the explicit formulas for the solutions and their associated thermodynamic quantities to the appendix \ref{app:background}. Remarkably, we notice that the black hole is electrically charged even in the absence of a chemical potential --a peculiar consequence of the presence of $\lambda_1$ and $\lambda_3$ (see Eq. \eqref{eq:ChargeDens}). Actually the black hole solution takes the simple form of the electrically neutral dyonic black hole, but with a non vanishing electric field. For $\mu=0$ the metric and gauge field are
\begin{eqnarray}
f(r)&=&1-\frac{B^2+4r_0^4}{4r_0^4r^3}+\frac{B^2}{4 r_0^4r^4 }\,,\qquad C(r) = 1\,,\\
a(r) &=& (4 \lambda_1-\lambda_3)\left(\frac{20 r_0^4- 3 B^2}{20 r r_0^5}-\frac{ \left(B^2+4 r_0^4\right)}{4 r^4 r_0^5}+\frac{2 B^2}{5 r^5 r_0^5}\right)B\,. \label{eq:Sola0}
\end{eqnarray}

We now proceed to study the thermodynamic properties of the state in the dual theory. The temperature of the dual theory equals to the Hawking temperature of the black hole, and can be obtained by continuing the geometry to Euclidean signature and imposing regularity at the horizon
\begin{equation}
T = \frac{r_0}{4\pi}f'(1)\,, \label{eq:T}
\end{equation}
notice that $r_0$ is a function of $\mu,B$ and $T$, which can be obtained by solving Eq. \eqref{eq:T}, also notice that the underlining conformal invariance implies that $r_0=g(B/T^2,\mu/T)T$.
The energy and charge densities are computed following the standard AdS/CFT dictionary and applying holographic renormalization methods (more details can be found in Appendix \ref{app:LCT}). We obtain the following expressions \begin{equation}
\varepsilon= 2(P-MB) = \frac{2c_T r_0^3}{3 }\lim_{r\to\infty}r^4f'(r) \quad,\quad \rho =8 c_TB  (2 \lambda_1+\lambda_3-\lambda_0)+ r_0c_T\lim_{r\to\infty}r^2a'(r)\,,\label{eq:parameters}
\end{equation}
where $c_T=L^2/2\kappa^2$ \footnote{If the dual is a large-$N$ gauge theory, $c_T\sim N^{3/2}$ \cite{Herzog:2007ij} roughly counts the number of microscopic degrees of freedom.}, $P$ is the pressure and $M$ is the magnetization $M=\frac{\partial P}{\partial B}$. The relation between the energy density and $P-MB$ follows from conformal invariance of the theory, but we have checked that it is satisfied by explicitly computing the expectation value of the stress tensor.

The entropy density is normally defined as the area of the black hole in Planck units, however, in the presence of higher derivative terms, there are additional contributions that can be computed using Wald's formula for the entropy \cite{WaldEntropy94}. In particular, for a static background the formula reads
\begin{equation}
S=\frac{2 \pi}{\kappa^2}\int\limits_{\Sigma}Q^{ABCD}\nabla_A\chi_{_B}\nabla_C\chi_{_D} \sqrt{\sigma}dx^{2}\equiv\int\limits_{\Sigma}s dx^{2},
\end{equation}
with $Q^{ABCD} = -\frac{\partial \mathcal{L}}{\partial R_{ABCD}}$, $\chi$ the killing field generating the isometry of the horizon and $\sqrt{\sigma}$ being the determinant of the induced metric on the horizon. Upon evaluating the above, we get that total entropy density takes the form
\begin{equation}\label{eq:EntrDens}
s=4\pi c_T \frac{d\text{Vol}_{2d}}{L^2}-4\pi c_T\lambda_1L^2\, ^\star F^{tr}F^{tr}=4\pi c_T\left(r_0^2C(1)+ Ba'(1)\frac{4\lambda_1}{r_0}\right).
\end{equation}
As can be seen in Eq. \eqref{eq:EntrDens}, the usual formula formula for the entropy as the area of the horizon gets modify by the $\lambda_1$ term. Actually the correction is proportional to the product of the magnetic field and electric flux evaluated at the horizon.

\begin{figure}[t]
\centering
\includegraphics[width=.5\textwidth]{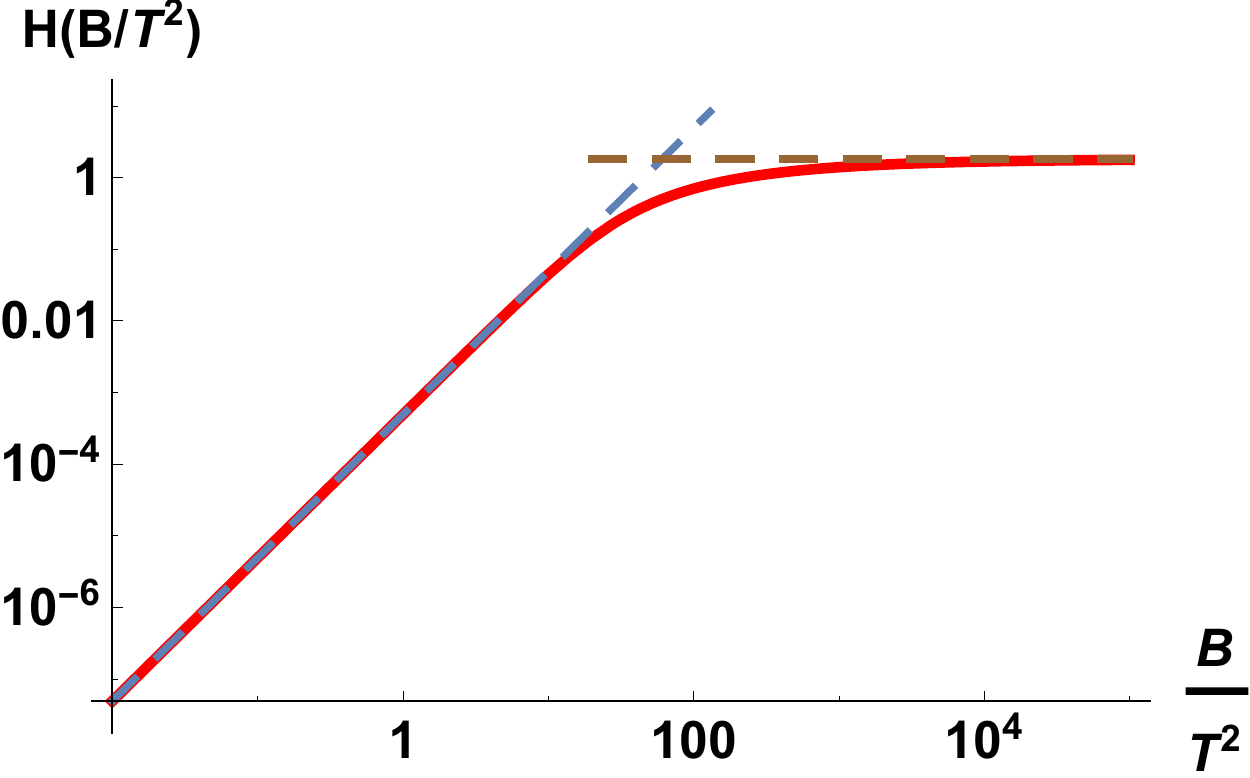}
\caption{Temperature dependence of $H$ in \eqref{eq:ChargeDensMuZero} as a function of the dimensionless parameter $B/T^2$. $H$ interpolates between a power-law $\sim (B/T^2)^2$ at high temperatures $B/T^2\ll 1$ and a constant at low temperatures  $B/T^2\gg 1$. Dashed lines show the asymptotic behavior.}\label{fig:Density}
\end{figure}

Particularizing the formulas above for the $\mu=0$ geometry \eqref{eq:Sola0}, the temperature and entropy density take the same form as in the dyonic black hole
\begin{eqnarray}
T = \frac{12 r_0^4-B^2}{16 \pi  r_0^3}\,,\qquad s = 4\pi c_T r_0^2\,.\label{eq:Ts}
\end{eqnarray}
We remark here that the electric flux at the horizon is $O(\lambda)$, so the correction to the entropy density vanishes to leading order but there can be subleading corrections that we have not computed. The charge density is non-zero, and takes the form
\begin{equation}
\rho_{_B}=c_T (4 \lambda_1-\lambda_3)H\left(\frac{B}{T^2}\right)B +c_T(12 \lambda_1+9 \lambda_3-8\lambda_0)B, \label{eq:ChargeDensMuZero}
\end{equation}
where the dimensionless function $H$ is represented in Figure \ref{fig:Density}. In the plot we observe two well defined asymptotic regions at high and low temperatures. At high temperatures the contribution proportional to $H$ goes to zero and the density is determined by the second term in \eqref{eq:ChargeDensMuZero}. At low temperatures  $H$ goes to a constant and both terms in the density contribute at the same order. In both cases the density has a linear dependence with $B$ at leading order , but different proportionality constant.

The energy density, pressure and magnetization of the system take the form
\begin{equation}\label{eq:epPM}
\varepsilon  =  c_T\frac{ \left(B^2+4 r_0^4\right)}{2 r_0}\,,\qquad P = c_T\frac{ \left(4 r_0^4-3B^2\right)}{4 r_0}\,, \qquad M = -c_T\frac{B}{r_0}\,.
\end{equation}
As can be checked from Eqs.~\eqref{eq:epPM} the system obeys the condition $T^\mu_\mu=0$ coming from conformal invariance.

\section{Transport coefficients}\label{sec:transp}

To obtain the transport coefficients, we study  linear response around the equilibrium state described by the black hole Eq. \eqref{eq:Ansatz}. 
Linear response in the context of holography corresponds to solving the problem of small perturbations propagating on such black hole geometry. We introduce a small parameter $\epsilon\ll 1$ that determines the amplitude of the perturbations and expand to linear order. The form of the metric and gauge field is
\begin{align}
\nonumber ds_\epsilon^2&=ds_{(0)}^2+\epsilon\,L^2 r_0^2r^2 C(r)\left[
h_x(r,t)\left(dx^2-dy^2\right) +2h_y(r,t) dxdy +2w_j(r,t) dtdx^j\right] +\mathcal{O}(\epsilon^2),\\
A&=a(r)dt+B\, x \,dy + \epsilon b_j(r,t) dx^j+\mathcal{O}(\epsilon^2).
\end{align}
Where the perturbations have been classified under their transformation properties under the SO(2) rotational invariance of the boundary theory. The terms $h_x$, $h_y$ are tensor modes that can be used to compute viscosities, while the terms $w_j$, $b_j$ are vector modes that are related to the conductivities. The remaining components form the scalar sector, that contains the bulk viscosity. Since the dual theory is conformal, the bulk viscosity vanishes $\zeta=0$, we have confirmed this result by explicit calculation but we do not show it here, as it is straightforward but quite lengthy, and not particularly interesting since no other transport coefficients belong to this sector.

\subsection{Viscosities}
\begin{figure}[t]
\centering
\includegraphics[width=0.5\textwidth]{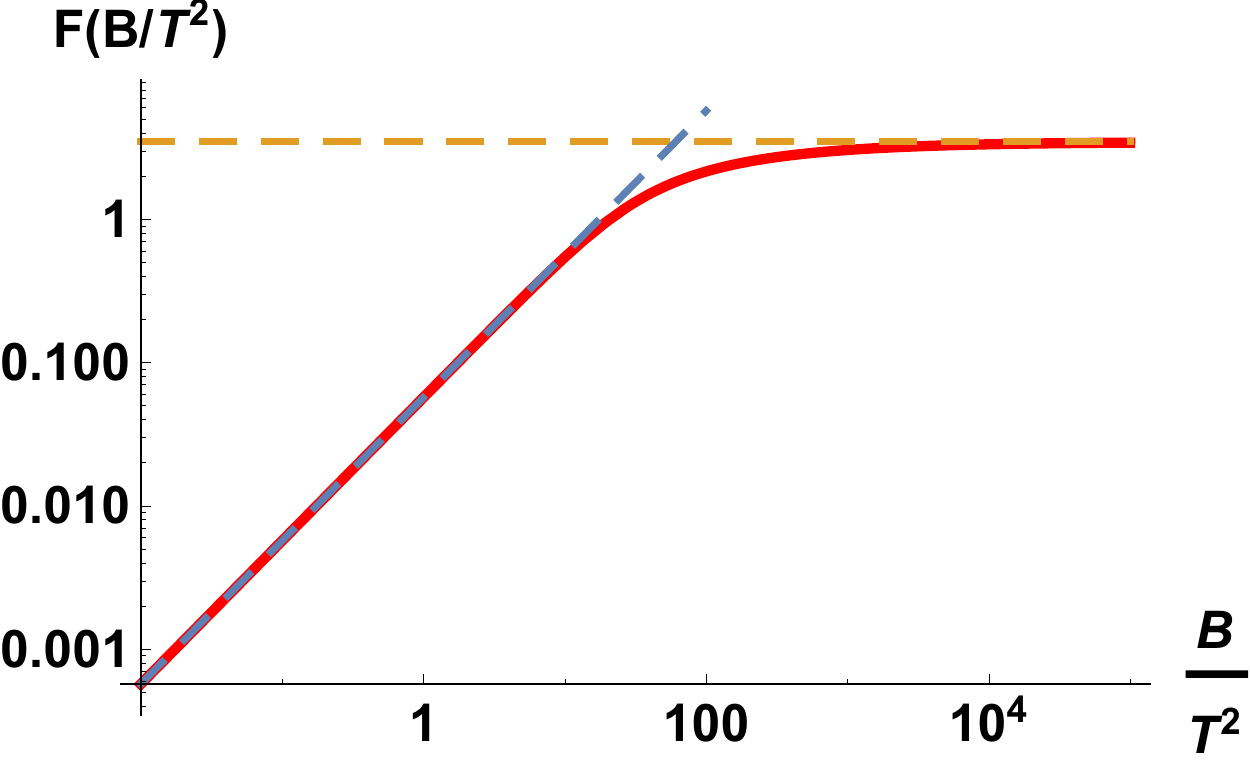}
\caption{Hall viscosity ($\eta_{_H}\sim F\left(\frac{B}{T^2}\right)B$) at zero chemical potential as a function of $B/T^2$ (red continuous line). Dashed lines show the asymptotic power laws. In particular blue dashed line corresponds to  $F= \frac{9B}{16\pi^2T^2}$, on the other hand orange dashed line corresponds with the fitting  $F = 2\sqrt{3}$.}\label{fig:HallDimensionless}
\end{figure}

Since the main goal of this work is to investigate the Hall viscosity in a strongly couple magnetized plasma, we begin discussing first the perturbations in the tensor sector, that encodes this response coefficient. This sector is constituted by the fluctuations $h_x(r,t),h_y(r,t)$ and responsible for the shear and Hall viscosities. In particular, after Fourier transforming the fields $h_i\to h_{i}(r) e^{-i \omega t}$,  the equations of motions can be written as follows
\begin{eqnarray}\label{eq:TensorPert}
\left(r^4 F_1(r) f(r)h_k'(r)\right)'+i\omega  \epsilon_{kj}F_2'(r)h_j(r) + \omega ^2 F_3(r) h_k(r)=0\,,
\end{eqnarray}
where
\begin{eqnarray}
F_1 &=& C(r)+ (2 \lambda_1-\lambda_3) \frac{2 B a'(r)}{r^2 r_0^3}\,,\\
F_2 &=& \frac{1 }{r_0^3}(4 \lambda_1-\lambda_3)\left(\frac{B^2 }{r^2 r_0^2 C(r)}- r^2 C(r) a'(r)^2\right)\,,\\
F_3 &=& \frac{ C(r)}{r_0^2 f(r)} + (2\lambda_1-\lambda_3)  \frac{2B a'(r)}{r_0^5r^2f(r) }\,.
\end{eqnarray}
 As the holographic dictionary establishes, we must find solutions satisfying an infalling boundary condition, which we guarantee by redefining the fields 
\begin{equation}
h_k(r)=f(r)^{- \frac{i\omega}{4\pi T}}p_k(r)\,,
\end{equation}
and then requiring regularity at the horizon for $p_k(r)$. Given the form of the Kubo formula Eq. \eqref{eq:TensorKubo}, it is only necessary the knowledge of the fields up to linear order in the frequency $\omega$. Therefore, we do a perturbative expansion in frequency such that  $p_k(r) = p_k^0(r)+ \frac{\omega}{4\pi T}\,p_k^1(r)$. After doing so, the equations of motion read
\begin{equation}
\left(r^4 F_1(r) f(r)p_k^{s\prime}(r)\right)'=S_k^s(r),
\end{equation}
with
\begin{equation}
S_k^0 = 0\,,\quad S_k^1 = i\frac{r^4 f'(r) }{f(r)}F_1(r)p_k^{0'}(r)+ i\left(\frac{r^4 f'(r) }{f(r)}F_1(r)p_k^{0}(r)\right)'-4i\pi TF_2'(r)\epsilon_{kj} p_j^0(r)\,.
\end{equation}
A set of two linearly independent solutions can be constructed imposing regularity at the horizon, and considering independent boundary values $h_y^0$ and $h^0_x$ which are dual to sources for the stress tensor components $T^{xy},T^{xx}-T^{yy}$. The system can be solved in terms of the background solutions without doing any additional approximations, the result being
\begin{equation}\label{eq:gk}
p_k = h_k^0 + \frac{\omega}{4\pi T} \int_{\infty }^r \frac{g_k(x)-g_k(1)}{x^4 f(x)F_1(x)} \, dx\,,\quad g_k =  i\frac{r^4 f'(r) }{f(r)}F_1(r)h_k^{0}-4i\pi TF_2(r)\epsilon_{kj} h_j^0\,.
\end{equation}
Which, after being plugged in Eq. \eqref{eq:OPFstress} and combined with the Kubo relations Eq. \eqref{eq:TensorKubo}, gives the values for the viscosities of the model 
\begin{align}\label{eq:viscosities}
\eta&= \left(\frac{s}{4\pi}+\frac{2c_T    \lambda_3 }{ r_0}  Ba'(1)\right)\approx c_Tr_0^2\left[1-(20 \lambda_1-3 \lambda_3)\frac{\mu B   }{3r_0^3}\right] +\mathcal{O}(\lambda^2)\\
\eta_{_H}&= (4   \lambda_1-\lambda_3)\left(C(1) a'(1)^2-\frac{B^2}{C(1) r_0^2}\right)\approx  c_T(4 \lambda_1-\lambda_3)r_0^{-2} \left((r_0\mu) ^2-B^2\right)+\mathcal{O}(\lambda^2)
\end{align}
Note that the Hall viscosity is not zero even at $\mu=0$ as long as the magnetic field does not vanish, 
\begin{equation}
\eta_{_H}=c_{T}(\lambda_3-4 \lambda_1) F\left(\frac{B}{T^2}\right)B\label{eq:HallDimensionless}\,,
\end{equation}
where, due to the underlying conformal invariance of the system, the function $F$ depends only on the dimensionaless combination $B/T^2$. In Fig. \ref{fig:HallDimensionless} we show the dependence of $F$ as a function of the ratio $B/T^2$, from which we conclude that for small values of $B/T^2$ the Hall viscosity grows quadratically with $B$, while for large values the Hall viscosity grows linearly with the magnetic field and becomes independent of the temperature. 

The shear viscosity at zero chemical potential satisfies the Kovtun-Son-Starinets (KSS) formula \cite{Kovtun:2004de}, but when $\mu\neq 0$, higher derivative corrections change the viscosity to entropy ratio, with the sign of the correction depending on the details of the model. In particular, the correction to the KSS formula is proportional to the coefficient $\lambda_3$, while the correction to the entropy density, Eq. \eqref{eq:EntrDens}, was proportional to $\lambda_1$.

\subsection{Conductivities}\label{sec:conductivities}

\begin{figure}[t]
 \centering
 \includegraphics[width=.5\textwidth]{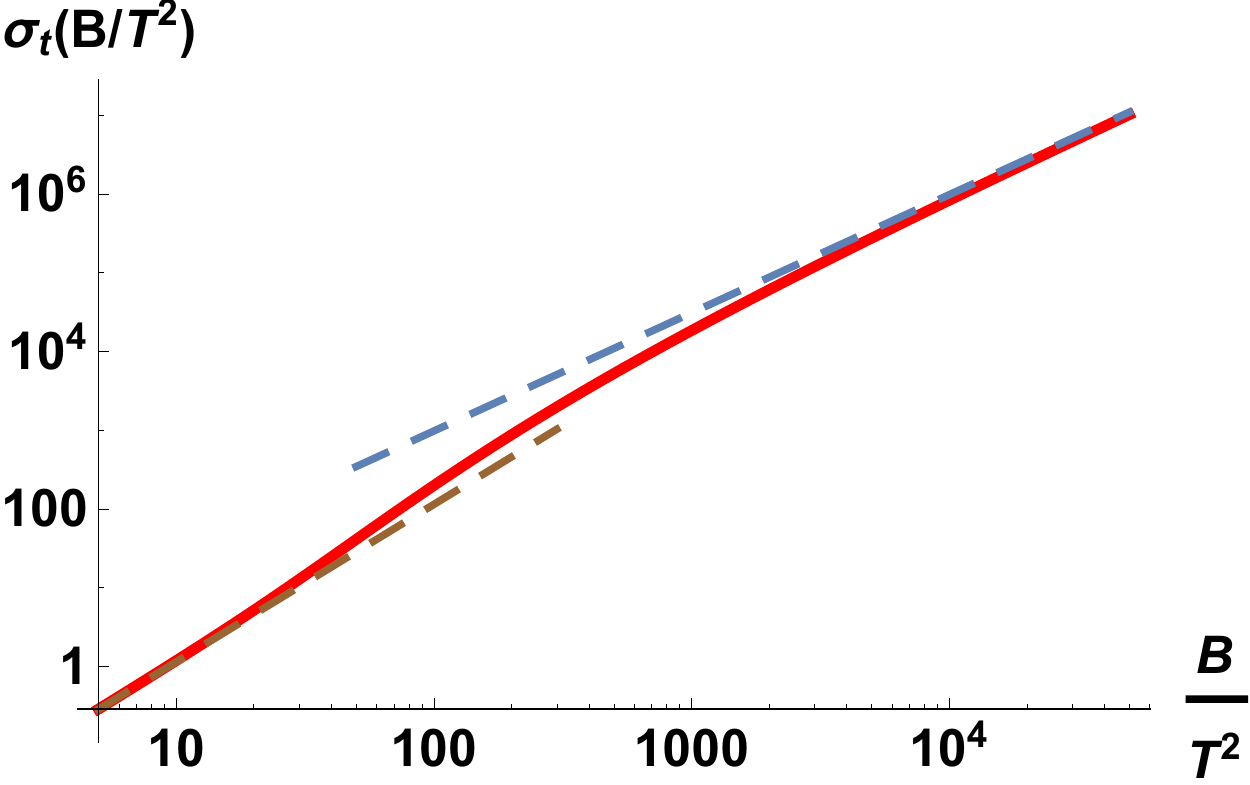}
 \caption{Function $\sigma_t$ in \eqref{eq:HallGeneral} as a function of $B/T^2$ (continuous red line). Brown dashed line shows the high $T$ behaviour $\sigma_t = B^2/T^4$. Blue dashed line represents  $\sigma_t = \left(B/T^2\right)^{3/2}$.}\label{fig:Sigma}
 \end{figure}

This sector is given by the fields $w_i(r,t),b_i(r,t)$, that are responsible for the charge and thermal conductivities.  The analysis is similar to the tensor modes, but the system of equations remains higher order in derivatives. In order to avoid the issue of having to solve higher order derivative equations, we do a perturbative expansion of the fluctuations in the couplings $\lambda_1\sim \lambda_3\sim \delta\ll 1$, as well as in the frequency $\omega/T\ll 1$. The fluctuations $\Phi_i=(w_i,b_i)$ are expanded in the following way
\begin{equation}
\Phi_i = \phi_i^{00}(r)+\delta \phi_i^{01}+\frac{\omega}{4\pi T}\left(\phi_i^{10}+\delta \phi_i^{11} \right) +\text{higher order terms},
\end{equation}
where the first upper index refers to the order in the frequency and the second to the order in the higher derivative couplings. We have left an explicit factor of the parameter $\delta$ to help follow the expansion, it should be noted that at the end of the calculation its value will be fixed to $\delta=1$.
Using this expansion yields a set of second order equations, as all the higher derivative terms in the original equations are already of order $\delta$, and so they all contribute only to source terms in the expanded equations. So, technically, the higher derivative terms are turned to higher derivatives of lower (in $\delta$) order solutions present in source terms. After imposing ingoing boundary conditions
\begin{equation}
w_j(r)=f(r)^{1-\frac{i\omega}{4 \pi T}}v_{j},~b_k=f(r)^{-\frac{i\omega}{4 \pi T}}q_k(r)\,,
\end{equation}
the system of equations takes the form
\begin{eqnarray}
&\left(r^4 f^2(r)\partial_r v^{\alpha\beta}_j \right)'=S^{\alpha\beta}_{j}\,,\label{eq:vector1}\\
&\partial_r q^{\alpha\beta}_j+ \mu v^{\alpha\beta}_j = \tilde S^{\alpha\beta}_{j}\,,\label{eq:vector2}
\end{eqnarray}
where $S^{\alpha\beta},\tilde S^{\alpha\beta}$ are source terms, that can depend on all the lower order perturbations and their derivatives up to order 3, which we show in the appendix \ref{app:sources}.  In contrast to the tensor sector, the vector perturbations are mixed by the equations of motion and yield odd transport coefficients also for $\lambda_1=\lambda_3=0$ \cite{Hartnoll:2007ai}. On the technical level this means that the source terms $S_v,~S_q$ are in general quite complicated even within this perturbative scheme and their integration is difficult. However, setting $\mu=0$  significantly simplifies the system without trivializing the transport coefficients. Therefore we solve this sector at zero chemical potential, and understand our results as the leading contribution in an expansion for small $\mu/T$. The Eqs. \eqref{eq:vector1} and \eqref{eq:vector2} have the same form as the equations found in \cite{Hartnoll:2007ai}, therefore we followed the same strategy to find the solutions. Finally, after solving them we proceed to substitute their solutions into the definition for the one point functions Eqs. \eqref{eq:OPFcurrent} and \eqref{eq:OPFstress} to extract the two point functions.
In this sector the current-current and current-momentum correlators satisfy the relations predicted by hydrodynamics Eqs. \eqref{eq:JJJT}. The momentum-momentum correlator reads
\begin{equation}
 G_{TT}^{0i,0j}=  \left(16 \pi ^2c_T\frac{  r_0^4 T^2  }{B^2}i\omega+\frac{\varepsilon}{2}\right)\delta^{ij}-\frac{3\pi  c_T T}{20 B^2 r_0^3}(4 \lambda_1-\lambda_3) \left(B^2 + 4 r_0^4\right)\left(7 B^2+60 r_0^4\right) i \omega\epsilon^{ij}\,,
\end{equation}
from which the heat conductivity $\bar\kappa$ can be obtained
\begin{equation}
T\bar \kappa^{ij} =   16 \pi ^2c_T\frac{  r_0^4 T^2  }{B^2}\delta^{ij}-\frac{3\pi  c_T T}{20 B^2 r_0^3}(4 \lambda_1-\lambda_3) \left(B^2 + 4 r_0^4\right)\left(7 B^2+60 r_0^4\right)\epsilon^{ij}\,.
\end{equation}
In fact, by comparing the expression for the heat conductivity with Eq. \eqref{eq:GTT}, we can fix the values for the transport coefficients $\sigma_{_V}$ and $\bar\sigma_{_H}$, which read
\be
\sigma_{_V}=9c_T\frac{(r_0-\pi T)^2}{\pi^2 T^2},
\ee
and
\be
\begin{split}\label{eq:Holosigmat}
	&\bar\sigma_{_H}=(4\lambda_1-\lambda_3)\frac{3B^2 c_T}{20 r_0^4}\left[ \frac{16\pi r_0^3T\sigma_{_V}^2}{9B^2c_T^2}\frac{9 r_0-7\pi T}{r_0-\pi  T}+1\right]+ c_T(12 \lambda_1+9\lambda_3-8\lambda_0).
\end{split}
\ee
We find convenient rewrite the previous expression as
\begin{equation}
\bar\sigma_{_H}= c_T(4 \lambda_1-\lambda_3)\sigma_t\left(\frac{B}{T^2}\right) + c_T\left(6(4\lambda_1+\lambda_3)-8\lambda_0\right)\,.\label{eq:HallGeneral}
\end{equation}
In Fig. \ref{fig:Sigma} we show the  functional dependence of $\sigma_t$  at $\mu=0$. For small values of $B/T^2$  $\sigma_t$ grows quadratically, while at  large values  $\bar\sigma_{_H}\sim \left(B/T^2\right)^{3/2}$. In the limit $B/T^2\to 0$ $\sigma_t$ vanishes, but there is a nonzero contribution to $\bar{\sigma}_{_H}$, corresponding to an anomalous Hall conductivity. 
  
  \subsection{Anomalous Hall conductivity}\label{sec:anomalousHall}
  
 It follows from  Eq. \eqref{eq:HallGeneral} that our system will in general have a non-vanishing anomalous Hall conductivity.  
 \begin{equation}
  \sigma_H^{\rm an}=c_T\left(6(4\lambda_1+\lambda_3)-8\lambda_0\right)\,, \label{eq:HallAnomalous}
\end{equation}    
However, do notice that the value of the anomalous conductivity can be arbitrarily tuned by changing $\lambda_0$. From the point of view of the effective dual field theory the presence of this coupling corresponds to the addition of a Chern-Simons term (see action Eq. \eqref{eq:Action}). This modification only affects the definition of the field theory $U(1)$ current, e.g. Eq. \eqref{eq:OPFcurrent}, and consequently the charge density Eq. \eqref{eq:ChargeDensMuZero} and Hall conductivity \eqref{eq:HallGeneral}. Interestingly, upon setting $\lambda_0$ to the value such that the anomalous Hall conductivity vanishes $\sigma_H^{\rm an}=0$, i.e.
\begin{equation}\label{eq:noanomcond}
\lambda_0=\frac{3}{4}(4\lambda_1+\lambda_3),
\end{equation}
all odd transport coefficients and the charge density become proportional to the combination $(4\lambda_1-\lambda_3)$ at $\mu=0$. In particular, the charge density \eqref{eq:ChargeDensMuZero} reduces to
\begin{equation}
\rho= c_{T} (4 \lambda_1-\lambda_3)\frac{3 B\left(B^2-20 r_0^4\right) }{20 r_0^4}\,.
\end{equation}
Interestingly, at large values of the magnetic field, the Hall viscosity becomes proportional to the charge density
\begin{equation}
\frac{\eta_{_H}}{\rho}=\frac{5}{\sqrt{3}}+\mathcal{O}(T/\sqrt{B})\,.
\end{equation}
This is similar to what happens in Quantum Hall states, with the difference that in known cases, such as Laughlin states, the coefficient is a fractional number.

\section{Discussion}\label{sec:discussion}

We have succeded in producing a non-zero Hall viscosity induced by a magnetic field in a holographic model via the introduction of higher derivative terms in the gravity action. The Hall viscosity, as given in \eqref{eq:viscosities}, receives two kinds of contributions, one proportional to the electric flux at the horizon produced by the charge inside the black hole $a'(1)$ and another one independent of it. The charge inside the horizon has been associated with ``fractionalized'' or deconfined degrees of freedom \cite{Sachdev:2010um,Hartnoll:2011fn}, while other contributions to transport have been dubbed confined, ``mesonic'' or ``cohesive'' \cite{Hartnoll:2012ux}. Cohesive transport is present for instance in holographic superfluids \cite{Hartnoll:2008vx,Hartnoll:2008kx,Iqbal:2011bf}, where some of the charge is carried by fields outside the horizon.

The fact that Hall transport is produced in part by the dynamics of the theory outside the horizon strongly suggests that similar behavior will also be possible in a geometry without horizons, including holographic duals to gapped states. This opens up the possibility of studying systems resembling much more Quantum Hall states, or at least some relativistic version. It should be noted that there is no obvious quantization of the coefficients of higher derivative terms, so there is no reason to expect Hall transport coefficients will be quantized even in the gapped systems. For the Hall conductivity, it is uncertain whether this is because we have not identified the right unit of conductance, it is a feature of the large-$N$ limit, or there is no intrinsic quantization in strongly coupled relativistic systems, further work would be needed to explore this issue. 

It would be interesting to understand better the meaning of the higher derivative terms from the point of view of the dual theory. Their structure suggests that they could be related to contributions in vacuum to the three point function of the energy-momentum tensor with two currents that are time-reversal and parity breaking. This could correspond to some type of explicit breaking by terms appearing at higher orders in a derivative expansion of the effective theory, or maybe an anomaly, although these are not topological terms. A weak coupling calculation could shed light on some of these questions.

Let us now comment a bit on the properties of transport coefficients and thermodynamics in the deformed dyonic black hole solution. For simplicity, we will restrict to the case with no anomalous Hall conductivity \eqref{eq:noanomcond} and $\mu=0$, although expressions for the thermodynamic quantities and viscosities at nonzero $\mu$ and other values of the coefficient $\lambda_0$ can also be found in the appendices \ref{app:background}. Since the theory is conformal, scaling arguments determine the dependence on the temperature/magnetic field with coefficients that can depend non-trivially on the dimensionless ratio $B/T^2$. It is thus equivalent to discuss low and high temperatures or large and small magnetic fields. It should be mentioned that although we have found that the results at large magnetic field are consistent with hydrodynamics, they are valid only as long as we consider small enough spacetime derivatives of the thermodynamic variables such that terms appearing at higher orders in the constitutive relations are suppressed. 

At high temperatures $B/T^2\ll 1$ thermodynamics is dominated by neutral degrees of freedom and the system behaves like a weakly magnetized conformal plasma
\be
 \varepsilon \simeq \frac{128 \pi ^3}{27}c_T T^3\,, \qquad P \simeq\frac{\varepsilon}{2}\,,\qquad  M \simeq -\frac{3c_T }{4 \pi }\frac{B}{ T}.
\ee
At low temperatures $B/T^2\gg1 $ there is ``vacuum'' contribution to the energy density
\be
\varepsilon\simeq \Lambda_{_B}=\frac{2 \sqrt{2} B^{3/2}c_T}{3^{3/4}}\,, \qquad P\simeq -\Lambda_{_B}+T s_{_B}+O(\sqrt{B}),
\ee
where the entropy is proportional to the magnetic field
\be
s_{_B} = \frac{2 \pi  Bc_T}{\sqrt{3}}.
\ee
The magnetization is determined to leading order by the vacuum contribution $M\sim -\sqrt{B}$.

The charge density varies little between high and low temperatures, becoming a $2/5$ factor smaller at low temperatures
\be
\begin{split}
&\frac{B}{T^2}\ll 1, \ \ \rho \simeq  3 c_T( \lambda_3- 4\lambda_1) B,\\
&\frac{B}{T^2}\gg 1, \ \rho \simeq \frac{6}{5} c_T( \lambda_3- 4\lambda_1) B.
\end{split}
\ee
The Hall conductivity then remains finite, with a change given by the same factor. Charge transport then remains largely independent of the temperature. The Hall viscosity however is more sensitive to the temperature. Taking the ratio with respect to the charge density
\be
\begin{split}
&\frac{B}{T^2}\ll 1, \ \ \frac{\eta_H}{\rho} \simeq  \frac{3}{16\pi^2}\frac{B}{T^2},\\
&\frac{B}{T^2}\gg 1, \ \frac{\eta_H}{\rho}  \simeq \frac{5}{\sqrt{3}}.
\end{split}
\ee
In contrast to Hall charge transport, at high temperatures Hall viscous transport is strongly suppressed. Thermal and thermoelectric conductivities involve both charged and neutral degrees of freedom, so we expect them to be more sensitive to the particulars of the dyonic black hole geometry, such as the non-zero entropy density at zero temperature. It is worth noting that for small values of the higher derivative coefficients, the Hall thermal conductivity $\kappa_H$ and the Seebeck coefficient $S=-\vartheta_{xx}$ are enhanced
\be
\begin{split}
&\frac{B}{T^2}\ll 1, \ \ \frac{\kappa_H}{T} \simeq   -\frac{(4\pi )^6}{3^5}  \frac{ c_T T^4}{B^2}\frac{1}{(\lambda_3-4\lambda_1)},\ \ S\simeq \frac{64 \pi ^3}{27(\lambda_3-4\lambda_1)}\frac{T^2}{B},\\
&\frac{B}{T^2}\gg 1, \ \frac{\kappa_H}{T}  \simeq  -\frac{10c_T\pi^2}{9}\frac{1}{(\lambda_3-4\lambda_1 )},\ \ S\simeq \frac{5\pi }{3\sqrt{3} (\lambda_3-4\lambda_1 )}.
\end{split}
\ee
It would be interesting to check if a similar enhancement would happen in other holographic models with different background geometry.

\section*{Acknowledgments}

F. P-B and C. H. acknowledge Nordita institute for hospitality. This work has been partially supported by the Spanish grant MINECO-16-FPA2015-63667-P, the Ramon y Cajal fellowship RYC-2012-10370 and GRUPIN 18-174 research grant from Principado de Asturias.

\appendix

\section{Variational Principle and renormalization}\label{app:LCT}

As it generically happens, also in our case the higher derivative nature of the action \eqref{eq:Action} spoils the variational principle. So, in this appendix we study the possible boundary terms that can be added to 'regularize' the variational problem and to renormalize the theory. In order to do so, we first assume the space-time can be ADM decomposed as follows.
\begin{equation}
ds^2 = dr^2 + \gamma_{ij} dx^i dx^j \,,
\end{equation}
with the gauge condition $A_r=0$, and the epsilon tensor defined as $\epsilon_{rtxy}=-\sqrt{-\gamma}$. The non vanishing components of the Christoffel symbols (and the  extrinsic curvature) are
\begin{align}
-\Gamma^r_{ij} &=  K_{ij} = \frac 1 2 \dot{\gamma}_{ij} \, ,\\
\Gamma^i_{jr} &= K^i_j \,,  
\end{align}
and $\hat\Gamma^i_{jk}$ are three dimensional Christoffel symbols computed with $\gamma_{ij}$. Dot denotes differentiation respect $r$.  Another useful table of formulas is
\begin{eqnarray}
\dot{\hat\Gamma}^l_{\,ki} \,\,\,&=& D_k K^l_i + D_i K^l_k - D^l K_{ki} \,, \\
R^r\,_{irj} &=& -\dot{K}_{ij} + K_{il}K^l_j \,,\\
R^k\,_{rjr} &=& -\dot{K}^k_{j} - K^k_{l}K^l_j \,,\\
R^r\,_{ijk} &=& D_k K_{ij} - D_j K_{ik} \,,\\
R^l\,_{kri} &=& D_k K^l_i- D^lK_{ik} \,,\\
R^i\,_{jkl} &=& \hat{R}^i\,_{jkl} - K^i_k K_{jl} +  K^i_l K_{jk}\,,
\end{eqnarray}
with $D_i$ the three dimensional covariant derivative compatible with $\gamma_{ij}$.
Note that indices are now raised and lowered with $\gamma_{ij}$, e.g. $K=\gamma^{ij} K_{ij}$, and intrinsic three dimensional objects are  denoted with a hat, so $\hat{R}^i\,_{jkl}$ is the intrinsic three dimensional Riemann tensor on the $r=\textrm{const}$ surface. Finally the Ricci scalar is
\begin{equation}
R = \hat{R} - 2 \dot{K} - K^2 - K_{ij}K^{ij} \,.
\end{equation}

Now, as we write the action, it is useful to divide it up in three terms. The first one is the Einstein-Maxwell part with the usual Gibbons-Hawking term included
\begin{equation}
\label{eq:Sb1} 
\mathcal L_{EM} =\frac{1}{2\kappa^2}\left(\hat{R} + 2 \Lambda + K^2 -
K_{ij}K^{ij} - \frac 1 2 E_i E^i - \frac 1 4 \hat{F}_{ij}\hat{F}^{ij} \right) \, , 
\end{equation}
where $E_i=\dot{A}_i$. The contributions to the action parametrized by $\lambda_1$ and $\lambda_3$ are
\begin{align}
\nonumber 	\mathcal L_{1} = -\frac{2L^4}{\kappa^2}\epsilon^{ijk}\left( E_l\left[2E_iD_kK^l_j - \hat F_{ij}K^l_m K^m_k +E_i \hat F_l\,^s\left(\hat R^l\,_{sjk}\right.\right.\right. &
+\left.\left. 2K^l_{[k}K_{j]s}\right) + 2\hat F_{ij}\hat F^{ls}D_{[s} K_{l]k} \right]+\\
\label{eq:act1} 	&\left.- \hat F_{ij}E^l\dot K_{lk} \right) \,,\\
\nonumber 	\mathcal L_{3} = \frac{L^4}{\kappa^2}\epsilon^{ijk} \left( E_m \hat F_i\,^l\hat R^m\,_{jlk} + E_m \hat F_i\,^lK^m_k K_{jl} - E_m E_i D_j K^m_k \right. &
+ \hat  F_i\,^l\hat  F_j\,^s D_s K_{lk} + \hat F_{il}E_j K^l_s K^s_k +\\
\label{eq:act3}& -  \left.\hat F_i\,^l E_j \dot K_{lk} \right) \, . 
\end{align}
Taking variations of the action, the last terms in Eqs. \eqref{eq:act1} and \eqref{eq:act3} which are proportional to $\dot K_{lk}$ will produce a boundary contribution of the type $\sim\delta K_{ij}\sim\delta\dot\gamma_{ij}$, suggesting we can regularize the problem by  adding the boundary term
\begin{equation}
\delta S_{GH} =-\frac{1}{\kappa^2}\int d^3x\sqrt{\gamma} \epsilon^{ijk}\left(2\lambda_1 F_{ij}E^l - \lambda_3F_i\,^lE_j\right)K_{lk}\,.
\end{equation}
However, $\delta S_{GH}$ would cancel the terms proportional to $\delta K_{ij}$ at the price of introducing a new term proportional to $\delta E_i$. Therefore, Dirichlet boundary conditions can be fixed either to the metric or to the gauge field, but not simultaneously to both. This fact resemble the case in $AdS_5$ with the mixed gauge-gravitational Chern-Simons term \cite{Landsteiner:2011iq,Copetti:2017ywz,Copetti:2017cin}, where a Gibbons-Hawking like term can be added but nonetheless, the regularity of the variational problem is not resolved. From a practical point of view, and in the context that concerns us, adding or not $\delta S_{GH}$ does not affect the observables because the near boundary behaviour of the fields in an asymptotically locally $AdS$ space is such that this boundary term always vanishes, as we discuss below.

On top of having a regular variational problem, the on-shell action needs to be finite, and so we proceed to find the proper counterterm which removes possible singularities. However notice that the counterterm renormalizing Einstein-Maxwell theory has being previously computed \cite{Balasubramanian:1999re}, therefore the boundary action must have the following structure
\begin{equation}
S_{CT} = -\frac{L}{2\kappa^2}\int d^3x\sqrt{\gamma} \left(\frac{4}{L^2}-\hat R\right) + \delta S_{CT}\,
\end{equation}
where $\delta S_{CT}$ is the contribution necessary to renormalize  the parity odd terms in the action.
As we have applications to holography in mind we can however impose the
boundary condition that the metric has an asymptotically locally AdS expansion
of the form
\begin{equation}
\gamma_{ij} = \e^{2 r} \left( g^{(0)}_{ij} + e^{-r} g^{(1)}_{ij}   + \cdots\right) \,, \qquad  A_{i} =  A^{(0)}_{i} + e^{-r} A^{(1)}_{i}   + \cdots \,,
\end{equation}
by inspection it is possible to conclude that $\delta S_{GB}$, $\mathcal L_1$ and $\mathcal L_3$ vanish fast enough on space-times with such asymptotic behaviour. Therefore,
\begin{equation}
\delta S_{CT}=0\,.
\end{equation}

Having studied the counterterms it is straightforward to compute the holographic one point functions of the $U(1)$ current and the stress energy tensor which read 
\begin{align}
\label{eq:OPFstress}T^{ij} &= -\lim_{r\to\infty}L^2r^2\frac{\sqrt{-\gamma}}{\kappa^2}\left(K^{ij}-K\gamma^{ij}+\frac{2}{L}\gamma^{ij}\right),\\
\label{eq:OPFcurrent}J^i &=-\lim_{r\to\infty}\frac{\sqrt{-g}}{2\kappa^2}\left( L^2F^{ri}+H_1^{ri}+H_3^{ri}+4\lambda_0 L^2\epsilon^{ijk}F_{jk}\right).
\end{align}

\section{Equations of motion and solutions}\label{app:eqs}

The structure of the equations of motion is the following
\begin{align}
&R_{MN}-\frac{L^2 R + 3}{2L^2}g_{MN}-\frac{L^2}{2} F_{(M|P|}F_{N)}^{~P}+\frac{L^2}{4}g_{MN}F_{PQ}F^{PQ} + T_{1~MN}+ T_{3~MN} =0,\label{eq:einstein}\\
&\nabla_{M}\left[L^2F^{MP}+ H_1^{MP}+ H_3^{MP}\right]=0.\label{eq:maxwell}
\end{align}
where 
\begin{align}
T_1^{GS}&=\lambda_1L^4\left(2\nabla_A \left(\nabla_L F^{A(G}\,^\star{}F^{S) L} \right)+R^{B(G}_{~~PQ}F^{S)}_{~B}\,^\star{}F^{PQ}\right),\\
H_1^{MN}&=-2\lambda_1L^4F_{AB}\left[ R^{BA}_{~~PQ}\epsilon^{MNPQ}-R^{MN}_{~~PQ}\epsilon^{ABPQ}\right],\\
T_3^{GB}&=\lambda_3L^4\left(F^{(GN}F^{M}_{~A}\epsilon_{MNQP}R^{AQPB)}-\nabla_{L}\nabla_{R}\left[F_N^{~L}\epsilon^{MNR(B}F_{M}^{~G)}\right]\right),\\
H_3^{MN}&=4\lambda_3L^4\epsilon^{APQ[M|}R^{|N]}\,_{PBL}F_A^{~B},
\end{align}
above $^\star{}F^{S L} = \epsilon^{{S L}MN}F_{MN}$ stands for the Hodge dual of $F_{MN}$.
Note, that due to the Lagrangian being fourth order in derivatives the given equations are in principle third order.

\subsection{Background solutions}\label{app:background}

In this appendix we show the background solution for arbitrary chemical potential, but up linear order in the odd couplings $\lambda_1,\lambda_3$. The black hole Ansatz reads
\begin{equation}
\frac{ds^2}{L^2}=\frac{1}{r^2f(r)}dr^2+r_0^2 r^2\left(-f(r)dt^2+C(r)\left(dx^2+dy^2\right)\right)\,,\quad A=a(r)dt+B x dy\,.
\end{equation} 
After evaluating Eqs. \eqref{eq:einstein} and \eqref{eq:maxwell} on it, linearizing the equations and solving them, we obtain the following solution for the background metric and gauge field 

\begin{eqnarray}
f(r)&=& \frac{(r-1) \left(-B^2-\mu ^2r_0^2+4 r \left(r^2+r+1\right) r_0^4\right)}{4 r^4 r_0^4}+\\
&&\nonumber -\frac{B \mu  (8 \lambda_1+3 \lambda_3) \left(B^2 \left(r^4-1\right)+\mu ^2 \left((3-2 r) r^4-1\right) r_0^2\right)}{60 r^8 r_0^7}+\label{eq:Solf}\\
&&\frac{B \mu   (r-1) (4 \lambda_1-3 \lambda_3) \left(B^2  \left(r^4-2\right)+\mu ^2\left(r^4-2\right) r_0^2+4 r (r (r (2-3 r)+2)+2) r_0^4\right)}{24 r^8 r_0^7},\nonumber\\
C(r) &=& 1+\frac{B \mu (8 \lambda_1+3 \lambda_3)}{3 r^4 r_0^3},\label{eq:Solc}\\
a(r) &=& \mu\left(1-\frac{1}{r}\right) -\frac{B \left(\mu ^2 r_0^2 (68 \lambda_1+3 \lambda_3)+3 (4 \lambda_1-\lambda_3) \left(3 B^2-20 r_0^4\right)\right)}{60 r r_0^5} +\label{eq:Sola}\\
\nonumber && +\frac{B  \left(6 B^2(4 \lambda_1-\lambda_3)+\mu ^2 r_0^2 (32 \lambda_1-3 \lambda_3)\right)}{15 r^5 r_0^5}-\frac{B(4 \lambda_1-\lambda_3) \left(B^2 +\mu ^2 r_0^2+4 r_0^4\right)}{4 r^4 r_0^5}.
\end{eqnarray}
The knowledge of the background allow us to compute the temperature $T$, and the one point functions of the current and stress-energy tensor after evaluating \eqref{eq:parameters}
\begin{align}
T \approx&\frac{12 r_0^4-B^2-\mu ^2 r_0^2}{16 \pi  r_0^3}+\frac{\mu  B  \left(3 B^2 (28 \lambda_1+3 \lambda_3)+\mu ^2 r_0^2 (52 \lambda_1-3 \lambda_3)+60 r_0^4 (3 \lambda_3-4 \lambda_1)\right)}{480 \pi  r_0^6}\label{eq:temp}\\
\rho  \approx& c_Tr_0\mu +  \left( \frac{3B^2}{20r_0^4} (4 \lambda_1-\lambda_3)+\frac{\mu ^2}{60r_0^2}  (68 \lambda_1+ \lambda_3)+3  (4 \lambda_1+3 \lambda_3)-8\lambda_0\right)c_TB .\label{eq:ChargeDens}\\
\varepsilon  \approx&  \frac{c_T \left(B^2+4 r_0^4+\mu ^2 r_0\right)}{2 r_0}+\frac{ c_T \mu  B\left(5 \left(B^2-12 r_0^4\right) (4 \lambda_1-3 \lambda_3)+\mu ^2 r_0 (52 \lambda_1-3 \lambda_3)\right)}{60 r_0^4},\\
M \approx& c_T\left(-\frac{B }{r_0}-\frac{\mu ^2 (r_0-1)}{2 B}\right)-\frac{4 c_T \mu  \left(6 \left(5 \lambda _0-16 \lambda _1-6 \lambda _3\right)
   r_0+\pi  \left(68 \lambda _1+3 \lambda _3\right) T\right)}{15 r_0}+\mathcal{O}(\mu^3),
\end{align}
and the pressure satisfy the relation $P = \frac{\varepsilon}{2} +B M\,$. 
Finally, we can also evaluate the entropy which reads
\begin{equation}
s\approx 4\pi c_Tr_0^2 \left(1-\frac{\mu  B (4 \lambda_1-3 \lambda_3)}{3 r_0^3}\right)+O\left(\lambda ^2\right)\label{eq:entropyOnShell}.
\end{equation}
Interestingly, although the density \eqref{eq:ChargeDens} at zero chemical potential is non-zero and a function of magnetic field \eqref{eq:ChargeDensMuZero}, the first order term in entropy density in \eqref{eq:entropyOnShell} is proportional to the chemical potential, and therefore vanishes when $\mu=0$. So, our holographic action describes a theory in which at zero chemical potential some charge density is induced by the presence of magnetic field, but this charge density does not contribute to the total entropy.

\section{Sources for the perturbative equations in the vectors sector }\label{app:sources}
In this appendix we show the explicit form of the sources introduced in Sec. \ref{sec:conductivities}, for simplicity we have set $\mu=0$ and evaluated the previous order solutions in some of them. 
\begin{align}
S_j^{00}&=0\\
\tilde S_j^{00}&=0\\
S_j^{10}&=-\frac{i \left(B^4 \left(r \left(r^2+r-8\right)+20\right)-8 B^2 r (5 r (r+1)+14) r_0^4+144 r^2 (r+1) r_0^8\right)}{(r-1) r^2 \left(B^2-4 r \left(r^2+r+1\right) r_0^4\right)^2}w^0_j \\
\tilde S_j^{10}&=\frac{i (r+1) \left(B^2-12 r_0^4\right) \left(B^2 (3 r-4)+12 r r_0^4\right)}{4 B r r_0 \left(B^2-4 r \left(r^2+r+1\right) r_0^4\right)}w^0_j \\
S_j^{01} &=0\\
\tilde S_j^{01} &=-\frac{B  (4\lambda_1-\lambda_3) \left(B^2 \left(3 r^4+20 r-40\right)-20 r \left(r^3-4\right) r_0^4\right)}{20 r^6 r_0^5}w^0_j\\
\nonumber S_j^{11}&=\frac{(4 \lambda_1-\lambda_3)\epsilon_{ji}w^0_i }{r^{11} r_0^{12} f_0(r)^2} \left(i r^{10} r_0^8 \left(B^2-12 r_0^4\right)  f_0'(r)^2+\right.\\
\nonumber&-\frac{1}{80} i r^5 r_0^4 \left(B^2-12 r_0^4\right)  f_0'(r) \left(B^2 \left(3 r^4+20 r-50\right)-240 r^4 r_0^4 f_0(r)+20 r \left(5 r^3+4\right) r_0^4\right)+\\
\nonumber&3 i r^8 r_0^8 \left(B^2-12 r_0^4\right) f_0(r)^2  -\frac{1}{4} i r^4 r_0^4 \left(B^2-12 r_0^4\right) f_0(r)  \left(B^2+12 r^4 r_0^4\right)\\
&\left. \frac{1}{320} i \left(B^2-12 r_0^4\right) \left(B^4 \left(7 r^5-12 r^4-20 r+32\right)-8 B^2 r \left((3 r-10) r^3+10\right) r_0^4-720 r^5 r_0^8\right)\right)\\
\nonumber \tilde S_j^{11}&=-\frac{B (4 \lambda_1-\lambda_3) v^{10}_i(r) \left(B^2 \left(3 r^4+20 r-40\right)-20 r \left(r^3-4\right) r_0^4\right)}{20 r^6 r_0^5}+\\
\nonumber& \frac{i w^0_i (r-1)}{320 B r^{10} f_0(r)r_0^9}(4 \lambda_1-\lambda_3)\left(B^6 \left(r \left(r \left(r \left(r \nonumber\left(3 r \left(r^2+r-6\right)+76\right)-12\right)-12\right)+48\right)-128\right)\right. +\\
\nonumber& -4 B^4 r (r (r (r (r (23 r (r+1)+26)+264)-56)-56)-176) r_0^4+\\
&\left. +48 B^2 r^2 (r (r (r (19 r (r+1)+70)+100)-20)-20) r_0^8-2880 r^5 \left(r^2+r-2\right) r_0^{12}\right)
\end{align}

\bibliographystyle{JHEP}

\bibliography{biblio}

\end{document}